# The Source-Normalized Impact per Paper (SNIP) is a valid and sophisticated indicator of journal citation impact

Henk F. Moed

*Senior scientific advisor at Elsevier, Amsterdam, and former professor of research assessment methodologies at Leiden University, the Netherlands.*

*Version 26 May 2010.*

The article "Scopus's Source Normalized Impact per Paper (SNIP) versus a Journal Impact Factor based on Fractional Counting of Citations", published by Loet Leydesdorff and Tobias Opthof (Leydesdorff and Opthof, 2010), denoted as L&O below, deals with a subject as important as the construction of 'field-normalized' indicators of the citation impact of scientific-scholarly journals.

Many authors have underlined that it is improper to make comparisons between citation counts generated in different research fields, because citation practices can vary significantly from one field to another. For instance, articles in biochemistry often contain over 50 cited references, while a typical mathematical paper has perhaps only 10. This difference explains why biochemical papers are cited so much more often than mathematical ones. Eugene Garfield's view on this is clear. "Evaluation studies using citation data must be very sensitive to all divisions, both subtle and gross, between areas of research; and when they are found, the study must properly compensate for disparities in citation potential (Garfield, 1979, p 249)". Advanced citation indicators should account for differences among subject fields in the frequency at which articles cite other documents, and the rapidity of maturing and decline of citation impact.

One of the solutions is applying the idea of source normalization, termed by Zitt and Small (2008) as "citing-side normalization". It can be carried out in several ways, but the base idea is that the actual citation rate of a set of target papers in a subject field is 'normalized' or 'divided' by a measure indicating the frequency at which articles in that field cite other documents. Moed (2010) developed a source normalized indicator of journal impact (SNIP) that was calculated for all journals indexed for Elsevier's Scopus, Since January 2010 this indicator is included in Scopus via de website www.journalmetrics.com, and also available at the site www.journalindicators.com hosted by the Centre for Science and Technology Studies at Leiden University, the institute at which SNIP was developed.

Both L&O's proposed indicator and SNIP are based on source normalization. Moreover, L&O adopt the idea underlying SNIP of defining a journal's subject field as the set of papers citing that journal. But they calculate ratios in a different way. While SNIP divides a journal's average number of cites per paper by the average length of reference lists in the journal's subject field – in fact, as explained below, only a selected part of those lists are taken into account – , L&O apply the principle of fractional citation counting (FCC) at the level of individual citing papers. If **$n$** is the number of cited references in a paper, each citation given by this paper counts as **$1/n$**. In short: while SNIP calculates a ratio of averages, L&O calculate an average of ratios. L&O claim that their measure is better than SNIP because it is "more simple and elegant", and enables one to test whether differences (e.g., between 2 journals) are statistically significant.

Below I clarify in Section 1 the relationship between SNIP and Elsevier's Scopus. Since L&O's description of SNIP is not complete, I indicate in Section 2 four key differences between SNIP and L&O's indicator, and argue why the former is more valid than the latter. Nevertheless, the idea of FCC deserves further exploration. In Section 3 I highlight two difficulties that arise if one attempts to apply this principle at the level of individual (citing) papers. Finally, in Section 4 I make concluding remarks.

**1. "Indicator of Scopus" means "indicator included in Scopus", not "developed by the Scopus Team".**

L&O denote SNIP as an "indicator of Scopus". Although it is included in Scopus as from January 2010 and calculated from Scopus data, it was developed and calculated by bibliometric researchers independently of Elsevier and the Scopus Team (Moed, 2010). The same is true for a second indicator launched at the same date in Scopus: SJR or Scimago Journal Rank (González-Pereira, Guerrero-Bote & Moya-Anegón, 2009).

**2. Key features of SNIP absent in O&L's measure**

SNIP is defined as the ratio of a journal's citation count per paper and the citation potential in its subject field. The base idea of citation potential is that the probability that a **$n$**-year old paper in a particular field is cited, is directly proportional to the frequency at which articles in the field cite other **$n$**-year old documents, in other words, proportional to the average number of **$n$**-year old cited references in the field's articles. I believe that SNIP measures citation potential more accurately than L&O's measure, in the following four respects.

i. SNIP's numerator, the journal citation count per paper, is defined as the average number of citations in a particular year, e.g., 2007 to 1-3 year old articles published in the journal. An appropriate definition of citation potential requires that the time windows in the measurement of citation counts and citation potential must be the same. Hence, citation potential is defined as the average number of 1-3 year old cited references in a journal's subject field. But in O&L's approach citation potential is seemingly based on the *total* number of cited references in a field's papers, regardless of whether they are published in the proper time window or not. In this way, journals in fields in which articles tend to have with long reference lists and citations a low immediacy, such as taxonomy, may be disadvantaged.

ii. SNIP's measurement of citation potential only takes into account cited references published in sources that are *indexed for Scopus*. For instance, citations to books are not included. To the extent that journal metrics is used to assess journals indexed for a database (in this case Scopus), it is appropriate to count only cited references actually published in indexed journals. In this way, one corrects for differences in *database coverage* across research fields. If not, the citation impact of indexed journals in fields in which database coverage is not as high as it is in (bio-)medical and physical sciences, such as mathematics, engineering, social sciences and humanities, would be systematically undervalued.

iii. A journal's subject field is indeed defined as the set of articles citing a journal's papers. But in the SNIP methodology, these articles do not necessarily have to cite 1-3 year old papers published in the journal, because this would introduce a bias in favor of articles citing recent materials above older documents. Therefore, a *ten*-year time window is applied: a journal's subject field consists of the articles citing at least one 1-10 year old paper published in the journal.

iv. A strong feature of SNIP is in my view that the range of values obtained BY SNIP is similar to that of the journal measure users are most familiar with, the journal impact factor published by Thomson-Reuters in its Journal Citation Reports. Only the highest SNIP values tend to be lower than the highest scores found in JCR. Further normalization of citation potential not explained in this reply is such that for half of a database' indexed journals have a relative citation potential below one. Hence, for half of the journals the SNIP ratio is higher than the value of a journal's average cites per article, and for the other half lower.

**3. Two problems related to fractional citation counting at the paper level**

If one would define citation potential in a proper way, as carried out in SNIP, it would still be possible to apply the key element in L&O's methodology, fractional citation counting (FCC) at the level of individual papers. If *r* denotes the number of a source article's 1-3 year old cited references published in indexed journals, each citation from this article to a particular target journal could be counted as *1/r*. But two major problems would arise, one of a statistical, and one of a theoretical nature.

i. As outlined above, to avoid a bias in favor of articles citing recent papers, a journal's subject field is defined as the set of articles citing at least one 1-10 year old paper published in the journal. A certain fraction of these articles will not have any 1-3 year old cited references; for these articles *r* equals zero. From the point of view of measuring citation potential, articles with zero cited references are as significant as those with any positive number. But how much would they contribute if one applies FCC? Is it not true that fractional citation counting would have to discard such articles, and hence generate a biased estimate of a subject field's citation potential?

ii. SNIP's denominator, citation potential, is a measure of how frequently papers in a journal's subfield cite on average other articles. Once it is calculated, it is the same for each (citing) article in that field, independent of the length of the citing article's reference list. But in the FCC approach each cited reference is weighted with 1/r (r=total number of cited references). What are the theoretical justifications and implications of this approach? Why should, – within one and the same subject field –, citations from papers with long reference lists count less than those from articles with short reference lists? In my view it makes sense to normalize at the field level, but less so at the level of individual articles. In any case, there is a theoretical issue at stake here that should be further debated.

**4. Concluding remarks**

A detailed discussion on the statistical tests O&L propose to apply to assess differences in citation impact among journals, and the base assumptions underlying such tests, goes beyond the scope of this reply. I refer to the work of Wolfgang Glanzel (2010) that shows that the suggestion that it would be impossible to develop such significance tests for indicators based on 'ratios of averages' rather than 'averages of ratios' is false.

Whether or not L&O's method is more elegant or simple than the SNIP method is mainly a matter of taste. In any case, validity should be the decisive criterion. In this reply I have argued why L&O's indicator is less valid than SNIP, and also highlighted two problems that would arise if one would apply fractional citation counting within the framework of the SNIP notion of citation potential. It would be interesting conducting more research into this FCC approach, examining not only statistical and theoretical aspects mentioned above, but also comparing its outcomes to those for SNIP, SJR and other journal impact indicators for a much larger set of journals than the five studied in O&L's paper.

**References**

Garfield, E. (1979). Citation Indexing. Its theory and application in science, technology and humanities. New York: Wiley.

González-Pereira, B., Guerrero-Bote, V.P. & Moya-Anegón, F. (2009). The SJR indicator: A new indicator of journals' scientific prestige. arXiv:0912.4141v1 [cs.DL].

Glänzel, W. (2010). On reliability and robustness of scientometrics indicators based on stochastic models. An evidence-based opinion paper. Journal of Informetrics, doi:10.1016/j.joi.2010.01.005.

Leydesdorff, L., & Opthof, T. (2010). Scopus' Source Normalized Impact per Paper (SNIP) versus a Journal Impact Factor based on Fractional Counting of Citations. arXiv:1004.3580v2 [cs.DL].

Moed, H.F. (2010). Measuring contextual citation impact of scientific journals. Journal of Informetrics, doi:10.1016/j.joi.2010.01.002.

Zitt, M., & Small, H. (2008). Modifying the journal impact factor by fractional citation weighting: The audience factor. Journal of the American Society for Information Science and Technology, 59, 1856-1860.